%% file: popwriteup.tex
\documentclass[aip, amsmath, amssymb,reprint,pop,floatfix]{revtex4-2}
\usepackage{amsfonts}
\usepackage[utf8]{inputenc}
\usepackage{graphicx} 

\usepackage{comment}
\usepackage[hidelinks]{hyperref}
\usepackage{xcolor}
\newcommand{\pdv}[2]{\frac{\partial #1}{\partial #2}}

\begin{document}

\title{Ponderomotive barriers in rotating mirror devices using static fields} 

\author{T. Rubin}
\email{trubin@princeton.edu}
\affiliation{Department of Astrophysical Sciences, Princeton University, Princeton, New Jersey 08540, USA}

\author{N. J. Fisch}
\affiliation{Department of Astrophysical Sciences, Princeton University, Princeton, New Jersey 08540, USA}
\date{\today}
\begin{abstract}
	 Particularly for aneutronic fusion schemes, it is advantageous to manipulate the fuel species differently from one another, as well as expel ash promptly. The ponderomotive effect can be used to selectively manipulate particles. It is commonly a result of particle-wave interactions and has a complex dependence on the particle charge and mass, enabling species-selectivity. If the plasma is rotating, e.g. due to  $\mathbf{E} \times \mathbf{B}$ motion, the ponderomotive effect can be generated using static (i.e., time-independent) perturbations to the electric and magnetic fields, which can be significantly cheaper to produce than time-dependent waves. This feature can be particularly useful in rotating mirror machines where mirror confinement can be enhanced by rotation, both through centrifugal confinement and additionally through a ponderomotive interaction with a static azimuthal perturbation. Some static perturbations generate a ponderomotive barrier, other perturbations can generate either a repulsive barrier or an attractive ponderomotive well which can be used to attract particles of a certain species while repelling another. The viability of each of these effects depends on the specifics of the rotation profile and temperature, and the resultant dispersion relation in the rotating plasma.
\end{abstract}
\maketitle

\section{Introduction}

Nuclear fusion of a proton-boron-11 fuel mixture, if possible, has a large upside potential. The fuel components are plentiful and easy to acquire, and require only a relatively simple isotope separation. The triple $\alpha$ reaction does not produce neutrons, causing no activation of reactor components. On the downside, the maximal cross-section of this reaction lies at $600$ keV, and steady state plasmas at these temperatures can lose significant power in Bremsstrahlung radiation. This was thought to be an insurmountable roadblock in the path of proton-boron-11 fusion~\cite{nevinsThermonuclearFusionRate2000,riderFundamentalLimitationsPlasma1995, riderGeneralCritiqueInertialelectrostatic1995,moreauPotentialityProtonboronFuel1977,levushPotentialityProtonboronFuel1982}. 

Recent work regarding the reaction cross-section~\cite{sikoraNewEvaluation$^112016} prompted re-examination of proton-boron-11 as a viable reaction for steady state fusion~\cite{weaver1973exotic,putvinskiFusionReactivityPB112019,ochsImprovingFeasibilityEconomical2022,caiStudyRequirementsP112022,eliezerIntroducingTwoTemperature2015,eliezerNovelFusionReactor2020,dapontaP11BMediumConfigurations2024,wangTransportAnalysisEHL22025,belloniMultiplicationProcessesHighDensity2022,horaRoadMapClean2017}. It appears that while power loss through photon radiation remains a challenge, is it possible to generate more fusion power than the expected radiation losses. In addition, some clever phase-space engineering could be used to further mitigate the severity of the radiative losses~\cite{qinAdvancedFuelFusion2024}.

By ``phase-space engineering'' we mean here any of the several methods of manipulating the particle distribution function either for the fusion fuel or the fusion ash. For the fuel, creating or maintaining non-Maxwellian features in the distribution is often advantageous, and \textit{resonant} wave-particle interaction can be exploited to heat the plasma~\cite{stixWavesPlasmas1992,dodinQuasilinearTheoryInhomogeneous2022,reimanSuppressionMagneticIslands1983,yoshiokaNumericalStudyMagnetic1984,hayeCrossMachineBenchmarking2006,reimanSuppressionTearingModes2018}, drive cross-field transport, or generate electric current parallel to the field~\cite{fischTheoryCurrentDrive1987}. The energetic fusion ash can be used to maintain non-Maxwellian features in the fuel population and be expelled at the same time~\cite{fischInteractionEnergeticAlpha1992,fischAlphaPowerChanneling1995,herrmannCoolingEnergeticParticles1997,ochsCouplingAlphaChanneling2015}.

\begin{figure*}
	\centering
	\includegraphics[width = \textwidth]{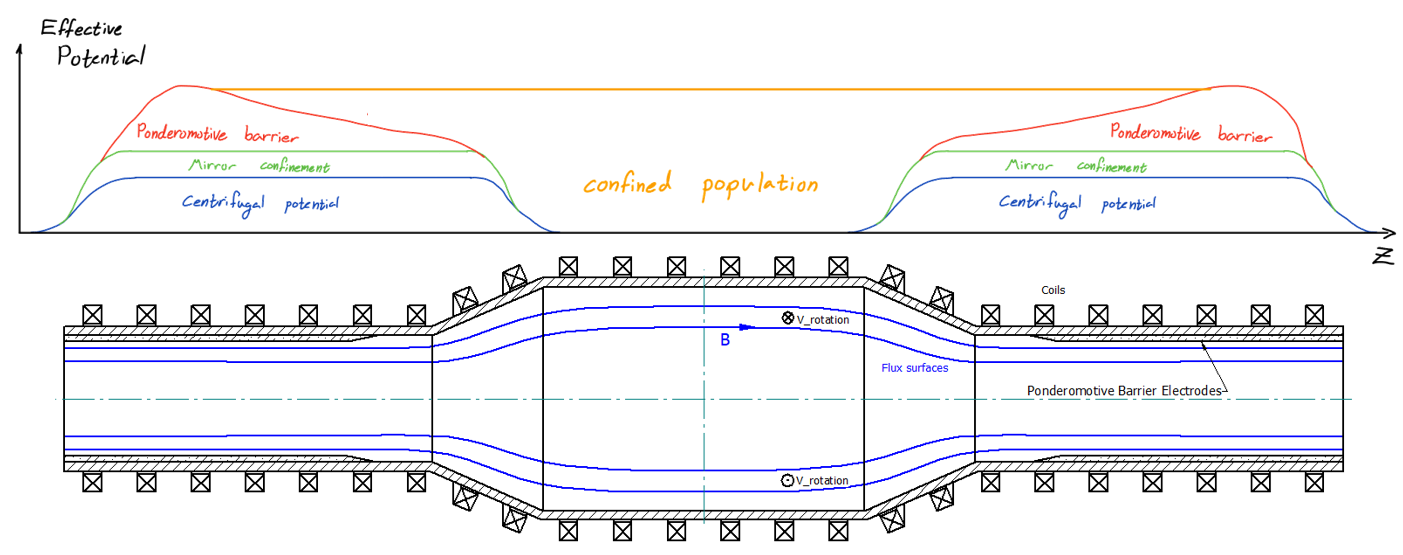}
	\caption{A schematic for the application of ponderomotive barriers in rotating mirror devices using static fields. In the Bottom figure: a sketch of rotating magnetic mirror machine, where the plasma rotating around the axis of the configuration. Magnetic flux surfaces are bent in the usual way to produce a magnetic mirror, with larger magnetic field at the mirror throats, and lower magnetic fields art the mid-plane. At the mirror throats, we add electrodes and coils to apply a static electromagnetic perturbation to generate a ponderomotive end-plug. In the Top figure, a schematic of the potential barriers affecting particles bouncing along the axis of the configuration, and the confined particle population.\label{fig: device sketch}}
\end{figure*}

The ponderomotive effect is one method of \textit{non-resonant}, or \textit{adiabatic} phase-space engineering. It can be used to generate an effective potential in the plasma, and is often the result of non-resonant particle-wave interaction~\cite{gaponov1958potential,lichtenbergRegularStochasticMotion1983, millerRFPluggingMultimirror2023, motzRadioFrequencyConfinementAcceleration1967,gormezanoReductionLossesOpenended1979,postMagneticMirrorApproach1987,dimontePonderomotivePseudopotentialGyroresonance1982}. Its magnitude is largest near resonance~\cite{dimontePonderomotivePseudopotentialGyroresonance1982}, and has a nonlinear dependence on particle charge and mass, gyroradius, as well as the field structure and polarization. The ponderomotive potential is flexible, and has been used to manipulate particles in for various purposes~\cite{chuNobelLectureManipulation1998, cohen-tannoudjiNobelLectureManipulating1998}, and appears in nature at various scales~\cite{guglielmiPonderomotiveUpwardAcceleration2001, shuklaDustGrainAcceleration2003}. In particular, it can generate potentials of either sign, i.e., can be repulsive or attractive~\cite{dodinApproximateIntegralsRadiofrequencydriven2005}. 

This effective potential in the particle path is of use not only in magnetic mirrors~\cite{burdakovMultiplemirrorTrapMilestones2016,fowlerNewSimplerWay2017,beeryPlasmaConfinementMoving2018,whiteCentrifugalParticleConfinement2018,ivanovGasdynamicTrapOverview2013,endrizziPhysicsBasisWisconsin2023,schwartzMCTrans0DModel2024}, but also in other devices using open magnetic field line configurations, such as isotope and mass separators~\cite{weibelSeparationIsotopes1980,gueroultOpportunitiesPlasmaSeparation2018,gueroultPlasmaFilteringTechniques2015,
gueroultPlasmaMassFiltering2014,gueroultDoubleWellMass2014,ohkawaBandGapIon2002,fettermanMagneticCentrifugalMass2011,oilerIncreasingEfficiencyPlasma2024,
hidekumaPreferentialRadiofrequencyPlugging1974,hiroeRadiofrequencyPreferentialPlugging1975,dolgolenkoSeparationMixturesChemical2017,timofeevTheoryPlasmaProcessing2014,litvakArchimedesPlasmaMass2003, voronaPossibilityReprocessingSpent2015}, which selectively confine different ion species based on their charge and mass. 

Other methods for phase-space engineering applying the ponderomotive effect with some resonance crossing have been proposed~\cite{dodinPonderomotiveBarrierMaxwell2004,dodinNonadiabaticPonderomotivePotentials2006}, where the sign-change of the ponderomotive interaction is employed to produce a diode-like potential in the plasma. 

\begin{figure}
	\centering
	\includegraphics[width=\columnwidth]{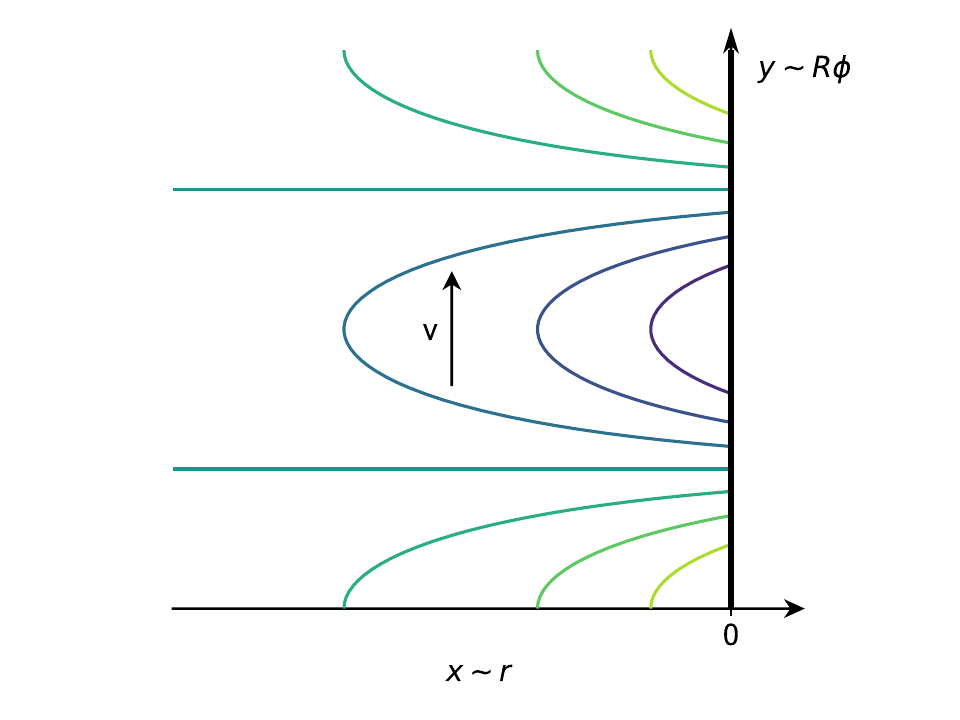}
	\caption{Illustration of plasma flow through a static perturbation. The arrow indicates the flow direction, parallel to the domain boundary, and the wave-vector of the perturbation has to have a component in the direction of the flow. The horizontal coordinate in this illustration can be either  the radial position for a realistic cylindrical case, or an equivalent for a slab analog. The vertical coordinate is the azimuthal coordinate or its analog.}\label{fig:flow through static perturbation}
\end{figure}

The oscillations in the particle dynamics generating the ponderomotive effect could be generated by plasma flowing through static perturbations~\cite{rubinGuidingCentreMotion2023, rubinMagnetostaticPonderomotivePotential2023,ochsCriticalRoleIsopotential2023a}, which are time-dependent waves in the moving frame. In Figure \ref{fig: device sketch}, we present a sketch of rotating magnetic mirror machine, with a sketch of the effective potentials affecting particles that bounce along the axis of the configuration. The sketch in Figure \ref{fig:flow through static perturbation} illustrates field configuration near the ponderomotive barrier electrodes, with plasma flowing parallel to the boundary of a domain, interacting with static electromagnetic field perturbation. This can be realized in a cylindrical geometry as illustrated in Figure \ref{fig: device sketch}, or simple slab models could be used for ease of calculations~\cite{ochsCriticalRoleIsopotential2023a,rubinFlowingPlasmaRearrangement2024a}. In this scenario, we propose a positive (repulsive) ponderomotive potential as an end-plug to the configuration, but an attractive potential could also be applied near the center of the device as well. 

Static perturbations are simpler and cheaper to implement over radio-frequency waves. Plasma rotation in magnetic mirrors and mass separators is useful on its own, and combining the two is a way to economically generate a useful effect.

In this work we assume that the plasma rotation has been arranged by separate means. Common methods to induce plasma rotation include concentric end-electrodes that are biased to produce a radial potential gradient, which may propagate from the electrodes into the plasma. The isorotation theorem~\cite{ferraroNonuniformRotationSun1937} provides for the near uniform rotation of the plasma on each drift surface. However, applying a perturbation at the ends of the device in order to generate a ponderomotive effect renders this approach more complicated, as drift surfaces would not remain axisymmetric in the presence of a non-axisymmetric perturbation. Some form of wave-induced rotation would be necessary.

This paper is organized as follows: In Sec. \ref{sec:adiabatic processes}, we discuss adiabatic processes in open field line configurations. We approach the subject from a Hamiltonian dynamics perspective, and describe the representation of the Hamiltonian in different coordinates. In Sec \ref{sec:static perturbations to flowing plasma}, we look into the possible perturbations consistent with plasma flowing over a set of static boundary conditions at the edge of the plasma. In Sec. \ref{sec: ponderomotive potentials}, we discuss the ponderomotive potentials generated by these perturbations.

\section{Adiabatic processes}\label{sec:adiabatic processes}

\textit{Non-resonant}, or \textit{adiabatic}, methods to manipulate the particle distribution function present themselves as ``quasipotential'' terms in the gyrocenter Hamiltonian. They belong in one of two categories; solvable dynamics generated by conserved adiabatic invariants and continuously changing frequencies, and the ponderomotive effect generated by oscillations. 

Adiabatic interactions keep the total particle energy, including potential energy, approximately constant. The particle energy can remain exactly constant, if there is no explicit time dependence in the Hamiltonian due to Noether's theorem~\cite{Noether1918}, because the energy is the conjugate ``momentum'' to the time coordinate. 

The Hamiltonian for a particle in electromagnetic fields that are derived from the electric potential $\Phi$ and the vector potential $\mathbf{A}$ is
\begin{gather}
	H = \frac{\left(\mathbf{p}-e\mathbf{A}(\mathbf{x},t)\right)^2}{2m}+e\Phi(\mathbf{x},t),\label{eq:Ham general}
\end{gather}
with $\mathbf{x}$ and $\mathbf{p}$ being the Cartesian spatial coordinates and their conjugate momentum, $e$ and $m$ are the particle charge and mass, and $t$ is time. This is equivalent to 
\begin{gather}
	H = \frac{1}{2}mv_x^2+\frac{1}{2}mv_y^2+\frac{1}{2}mv_z^2+e\Phi(\mathbf{x},t).\label{eq:cart energy}
\end{gather}
When looking at axisymmetric open magnetic field line configurations, with the vector potential having only a $\phi$ directed component, this Hamiltonian could also be written using cylindrical coordinates as
\begin{gather}    
    H = \frac{1}{2m}\left(p_z^2+p_r^2+\frac{(p_\phi-erA_\phi)^2}{r^2}\right)+e\Phi(r,z,t),
\end{gather}
using the $r$, $\phi$, $z$ coordinates. In an axisymmetric system, $H$ does not depend on $\phi$, and $p_\phi$ is a Noether invariant. Radial particle confinement is achieved due to the conservation of $p_\phi$, using Hamilton's equations,
\begin{gather}
	\dot \phi =\pdv{H}{p_\phi} = \frac{p_\phi-erA_\phi}{mr^2}.\label{eq:H cylindrical}
\end{gather}
With $A_\phi\approx \frac{1}{2}rB_z$ to leading order in $r$, and $B_z$ being the $z$ component of the magnetic field, for a constant $p_\phi$, the radial extent of motion is limited, i.e, particles perform helical motion around field lines and remain at a gyroradius distance from a field line, particle confinement in these configurations depends on the presence of a sufficient potential barrier in the path of the particle, larger than its parallel kinetic energy. Axial confinement is determined by a reflection along $z$, or $\dot z=p_z= 0$. For a fixed value of $H$ this means energy moving from the $1/2 m v_z^2$ term in equation (\ref{eq:cart energy}) or (\ref{eq:H cylindrical}) to any of the other terms. 

\begin{figure}
	\includegraphics[width = 0.6\columnwidth]{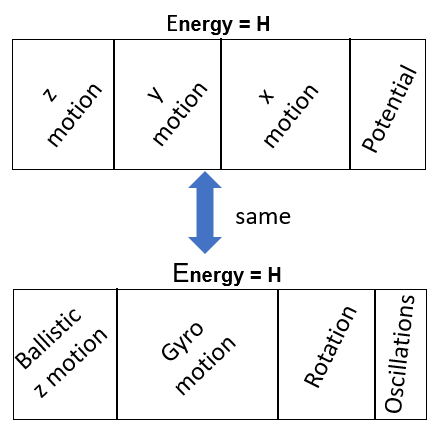}
	\caption{Illustration of the partition of the energy by direction of motion in a Cartesian system (top) or partition to action-angle coordinates (bottom).}\label{fig:Ham cartesian and action-angle}
\end{figure}

The Hamiltonian could also be expressed in terms of action-angle coordinates. For particle motion in axisymmetric electromagnetic fields, the particle position can be defined by the gyrophase $\theta$, and the magnetic moment $\mu=\frac{1}{2}m\Omega \rho^2$  (the first adiabatic invariant), with $\rho$ being the gyroradius and $\Omega$ a gyrofrequency, as well as the canonical angular momentum, and its conjugate phase, as well as the axial coordinate and its conjugate momentum. The Hamiltonian becomes
\begin{gather}
	H= \frac{p_z^2}{2m}+\Omega \mu + \omega_{rot}p_\phi+\Phi_z+\sum V_{m,n}e^{i( m\phi+n\theta)}, \label{eq:action angle Hamiltonian}
\end{gather}
where $\omega_{rot}$ is the rotation frequency around the axis of the configuration and $\Phi_z$ is an effective potential energy along z. The sum in the last terms describes the non-solvable, as well as the non-axisymmetric terms. $V_{m,n}$ could depend on $z$, $p_z$, $\mu$, and $p_\phi$. This partition of the energy is illustrated in Figure $\ref{fig:Ham cartesian and action-angle}$.

\begin{figure}
	\includegraphics[width = 0.6\columnwidth]{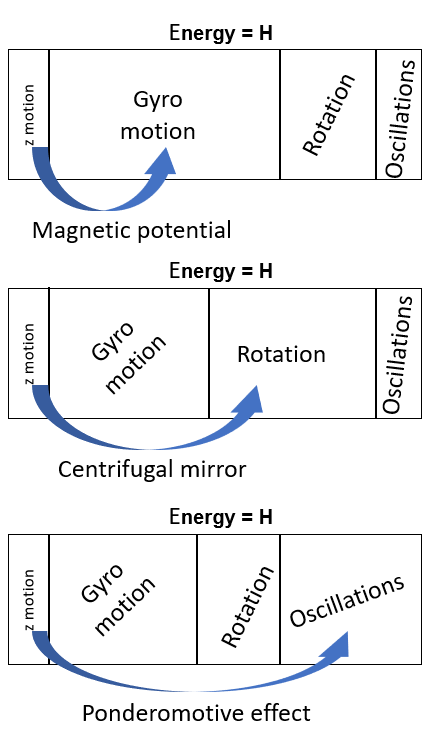}
	\caption{Illustration of the energy transfer from the axial ballistic motion into the other components of the Hamiltonian due to various processes.}\label{fig:Ham adiabatic processes}
\end{figure}

The solvable part of the Hamitonian consists of the first 4 terms in equation (\ref{eq:action angle Hamiltonian}). If the frequencies $\Omega$ and $\omega_{rot}$ or the potential energy $\Phi_z$ depend on $z$, these terms would be expressed as an effective potential due to Hamilton's equations. The effective potential barrier, measured from the point of minimum field, labeled with the index 0, to the point of maximum fields would be $(\Omega_{max}-\Omega_{0}) \mu +(\omega_{rot,max}-\omega_{rot,0}) p_\phi+ \Phi_z$. Figure \ref{fig:Ham adiabatic processes} illustrates energy transfer from the axial degree of freedom to the gyromotion degree of freedom (due to the dependence of $\Omega$ on z), or the rotation degree of freedom (due to the dependence of $\omega_{rot}$ on z), as well as the ponderomotive effect  (due to the dependence of $V_{m,n}$ on z).

Particles are axially confined if their parallel kinetic energy at minimum field $W_{\parallel 0}=\frac{p_{\parallel 0}^2}{2m}$ is smaller than the effective potential barrier. Often, we call the first term in the Hamiltonian the perpendicular kinetic energy $W_{\perp} = \Omega \mu $. The first two terms, $E_{kin}=\frac{p_\parallel^2}{2m}+\Omega \mu$ can be thought of as the kinetic energy. 

Unconfined particles escape the device, leaving a depleted region of phase space. The depleted region of phase space is determined by the diamagnetic effective potential~\cite{postMagneticMirrorApproach1987,ioffePLASMACONTAINMENTADIABATIC1970} and other effective potential terms such as the rotating mirror confinement for rotating mirrors~\cite{bekhtenevProblemsThermonuclearReactor1980,lehnertRotatingPlasmas1971}. 

\begin{figure}
	\centering
	\includegraphics[width=\columnwidth]{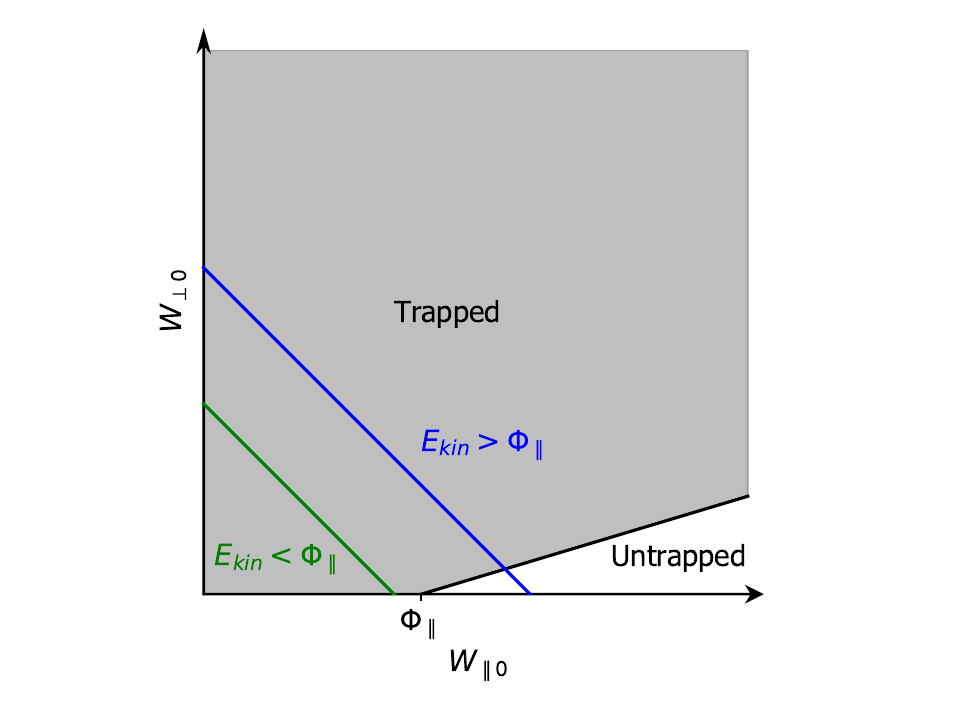}
	\caption{Phase space of particles in a magnetic mirror. Trapped particles satisfy $W_{\parallel 0 }<(R-1)W_{\perp 0}+\Phi_\parallel$, with $W_{\parallel 0 }$ beign the parallel kinetic energy at the minimum field, $W_{\perp 0 }$ being the perpendicular kinetic energy at the minimum field, $R = \Omega_{max}/\Omega_{0}$ being the mirror ratio, and $\Phi_\parallel$ being an effective potential barrier. The effective potential can be any combination of electric potential, centrifugal potential and ponderomotive potential. The untrapped region, ``loss cone'', is a nonisotropic feature preventing the distribution from relaxing to a Maxwellian, for $\Phi_\parallel$ comparable with the temperature.}\label{fig:phase space}
\end{figure}

The phase space of a simple magnetic mirror is presented in Figure \ref{fig:phase space}. The loss cone, which is an anisotropic feature, prevents the distribution function from relaxing to a Maxwellian, when $\Phi_z$ is comparable with the temperature~\cite{ochsElectronTailSuppression2024}. The value of $\Phi_z$ is not necessarily the same for all species. When it is an electrostatic potential, for example, it has opposite signs for ions and electrons.

The ponderomotive effect can be derived from oscillating terms in the Hamiltonian by applying a coordinate transformation~\cite{caryLieTransformPerturbation1981, depritCanonicalTransformationsDepending1969,dewarRenormalisedCanonicalPerturbation1976, zhaoLieseriesTransformationsApplications2025}, which is designed to remove the phase-dependence from the transformed Hamiltonian. This coordinate transformation is useful when there is a separation of time scales between the envelope $V_{m,n}$ changing and the gyromotion and rotation frequencies. In this case, the average action of the oscillating terms is the ponderomotive effect. The ponderomotive effect can act as an additional potential term to $\Phi_\parallel = \Phi_z+\Phi_{pond}$.

In magnetic mirror machines, the loss cone can be responsible for particle and energy loss both on the diffusion time scale and on faster time scales. On the diffusion time scale, pitch angle scattering of particles into the loss cone is a particle and energy loss mechanism in stable regular mirrors. On faster time scales, mirror instabilities can be triggered in the presence of a loss cone~\cite{kolmesLossconeStabilizationRotating2024,tranDriftcyclotronLossconeInstability2024}. The effect of the loss cone can be mitigated by increasing $\Phi_\parallel$, preferably for all species.  

In order to apply a ponderomotive potential barrier in a rotating plasma using static fields, we have to apply a static electromagnetic perturbation to a magnetic mirror configuration.

\section{Static perturbations to a flowing plasma}\label{sec:static perturbations to flowing plasma}
In recent works, we looked into applying perturbation to two different types of plasmas. In Ref~\onlinecite{rubinFlowingPlasmaRearrangement2024a} we investigated a simplified slab model and dense plasmas. In Refs~\onlinecite{rubinGuidingCentreMotion2023, rubinMagnetostaticPonderomotivePotential2023}, investigated a cylindrical geometry and a tenuous plasma.

\subsection{Slab geometry}
The slab system we consider has the same geometry described in Figure \ref{fig:flow through static perturbation}, with $x$ being the ``radial'' direction, $y$ being the ``azimuthal'' direction and $z$ being the ``axial'' direction. Plasma rotation around the device axis is analogous to flow in the $y$ direction. The domain of solution is the half volume $x<0$ in which the plasma resides.

The single particle picture in the slab is simple. Particles interacting with uniform crossed electric and magnetic fields perform uniform drift motion. Using the generating function $F=F(p_x,y,P_Y,\theta)$ 
\begin{gather}
	F = \frac{p_x^2}{2m\Omega}\cot{\theta}-p_x \frac{P_Y}{m\Omega}+y\left(P_Y-m\frac{E_0}{B_0}\right),
\end{gather}
for the canonical transformation of the $(x,y,p_x,p_y)$ coordinates to the $(\theta,Y,\mu,P_Y)$ coordinates such that the gyroradius $\rho = \sqrt{2\mu / m\Omega}$ and
\begin{align}
	x &= \rho \cos \theta + \frac{P_Y}{m\Omega}, \quad y = Y-\rho\sin\theta,\nonumber\\
	p_x &= m \Omega \rho \sin \theta,\quad p_y = P_Y-m\frac{E_0}{B_0} = P_Y+mv.\label{eq:slab action angle}
\end{align}
The variables of this transformation are understood as $Y$ denoting the $y$ coordinate of the gyrocenter, and $\mu$ being a measure of the gyroradius and the action conjugate to $\theta$, the gyrophase. The $x$ position of the gyrocenter is $P_Y/m\Omega$, with $P_Y$ being the conjugate momentum to $Y$.   

The Hamiltonian in equation (\ref{eq:Ham general}), with $\mathbf{A} = x B_0 \mathbf{e}_y$ and $\Phi = -x E_0$, i.e. with the fields $\mathbf{B}=B_0\mathbf{e}_z$ and $\mathbf{E}=E_0 \mathbf{e}_{x}$ is transformed into
\begin{gather}
	H = \frac{p_z^2}{2m}+\Omega \mu+P_Y v +\frac{1}{2}m v^2, \quad v = -\frac{E_0}{B_0}.
\end{gather}
With $\Omega$ being the cyclotron frequency, with opposite signs for ions and electrons, and the flow velocity $v$ being species-independent. The result here is that all plasma species flow together in these fields. 

Because the plasma flows together, the question of which perturbations are consistent with flowing plasma can be answered by a frame transformation, as was done in~\cite{ochsCriticalRoleIsopotential2023a,rubinFlowingPlasmaRearrangement2024a}. The procedure undertaken in these papers is to start with the dispersion relation of a stationary (not flowing) plasma. This is a moving frame moving with the plasma. Find the solutions to this dispersion, and the polarization of the electromagnetic fields in this frame, and perform a Lorentz boost into the lab frame, which is a frame moving with velocity $-v \mathbf{e}_y$ compared to the moving frame. 

The class of perturbations which interact with a flowing plasma in a way consistent with the ideas described above, is one with $\mathbf{k}\cdot\mathbf{v}\neq0$, i.e. a wave with some $k_y\neq 0$, with $k$ being the wave vector.

The Lorentz transformation in a flat spacetime of signature $-,+,+,+$ of a wave vector $\mathbf{k}$ and frequency $\omega$ to a frame moving with the plasma $\mathbf{k}',\ \omega'$ is
\begin{gather}
    \mathbf{k}' = \begin{pmatrix}
		k_x & \gamma (k_y-\beta \omega /c) & k_z
	\end{pmatrix},\label{eq:k trans}\\
	\omega'=\gamma(\omega-v k_y),
\end{gather}
with $\beta = v/c$ and $\gamma = \left(1-\beta^2\right)^{-1/2}$. That is, if the perturbation is time-independent in the lab (not primed) frame, 
\begin{gather}
	\omega=0,\\
    	\mathbf{k}' = \begin{pmatrix}
		k_x & \gamma k_y & k_z
	\end{pmatrix},\\
	\omega'=-\gamma v k_y.
\end{gather}

A perturbation applied at the $x=0$ plane would have the $k_y$ and $k_z$ components of the wave vector dictated by the boundary conditions, and the dispersion relation would determine the permissible values of $k_{x'}=k_{x}$ and the polarization. Even though it is tractable to consider any $k_z$, we elect to restrict ourselves to the case of $k_z=0$ for simplicity.

The dispersion relation for a simple uniform cold fluid plasma is given by~\cite{stixWavesPlasmas1992}
\begin{gather}
    \begin{pmatrix}
        S-N_{y'}^2-N_{z'}^2 & -iD+ N_{x'}N_{y'} & N_{x'}N_{z'}\\
        iD+N_{x'}N_{y'} & S-N_{x'}^2-N_{z'}^2 & N_{y'}N_{z'}\\
        N_{x'}N_{z'}& N_{y'} N_{z'}& P-N_{x'}^2 -N_{y'}^2
    \end{pmatrix} \mathbf{h} '    =0,
\end{gather}
with
\begin{gather}
    S = \frac{1}{2}(R+L), \ \ 
    D = \frac{1}{2}(R-L),\\
    R,L=1-\sum_s \frac{\omega_{ps}'^2}{\omega'(\omega'\pm\Omega_{s}')}, \ \ P = 1- \sum_s \frac{\omega_{ps}'^2}{\omega'^2}.
\end{gather} 
Where $\mathbf{N}'=\mathbf{k}'c/\omega'$ is the refractive index in the primed frame, $\mathbf{h}'$ is the electric field (complex) polarization vector, $\omega_{ps}'^2 = Z_s^2 e^2 n_{s}'/\epsilon_0 m_s$ is the plasma frequency of species $s$, $n_{s}'$ is its number density in the primed frame, and $\Omega_{s}' = Z_s e B_0' / m_s$ is its cyclotron frequency in the primed frame.

The solutions to this dispersion with $N_{z'}=0$ are the O wave and the X wave. 

\subsubsection{O wave}
Writing $k_y=k$, the dispersion of the O wave in the frame moving with the flow
\begin{gather}
    N_{x'}^2=P-\beta^{-2} ,\\
    k_{x'}= -i \sqrt{k^2+\sum_s \frac{\omega_{ps}'^2}{c^2}}=-ik\kappa_O,\\
    \kappa_O = \sqrt{1+\sum_s \frac{\omega_{ps}'^2}{c^2k^2}}.\label{eq:k o}
\end{gather}
The wave vector component in the $x'$ direction being imaginary renders this evanescent in this direction. The electric field polarization for the O wave is
\begin{gather}
	\mathbf{h}' = \mathbf{e}_{z'}\parallel \mathbf{B}' = \frac{B_0}{\gamma}\mathbf{e}_{z'}.
\end{gather}
The wave vector component in the x direction is dominated by the electron response, and by the triangle inequality,
\begin{gather}
	|k_{x'}|>\sqrt{\frac{\omega_{pe}'^2}{c^2}}=\sqrt{\frac{e^2n_e}{\gamma\epsilon_0 m_e c^2}}\approx \sqrt{\frac{n_e}{\gamma \ 10^{14} cm^{-3}}}18.8\ cm^{-1}.\label{eq:scale}
\end{gather}
From equation (\ref{eq:scale}) we see that the perturbation is evanescent on a short length scale, which is called the electron skin depth. This perturbation can be of relevance in the lab using tenuous plasma regimes, where electron density is a few orders of magnitude smaller than $10^{14}cm ^{-3}$, or using relativistic flow rates such that $\gamma \gg 1$.

The electromagnetic potential of this perturbation in the moving frame is, up to a phase
\begin{gather}
	\mathbf{A}_O' = -\frac{E_1'(z)}{\gamma k v} e^{k\kappa_Ox'}\sin(\gamma k (y'+vt'))\mathbf{e}_{z'},\\
\end{gather}
and the electromagnetic fields in the limit of $\pdv{E_1'}{z}=0$ are
\begin{gather}
	\mathbf{E}_O' = E_1' e^{k\kappa_Ox'}\cos(\gamma k (y'+vt'))\mathbf{e}_{z'},\\
	\mathbf{B}_O' = - \frac{E_1'}{\gamma v} e^{k\kappa_Ox'}\times\nonumber\\\left(\gamma\sin(\gamma k (y'+vt'))\mathbf{e}_{x'}+\kappa_O\cos(\gamma k (y'+vt'))\mathbf{e}_{y'}\right).
\end{gather}

A Lorentz boost of the wave electromagnetic fields back to the lab frame yields the time-independent perturbation is
\begin{align}
	&\mathbf{E}_O =0,\\
	&\mathbf{B}_O = B_1 e^{k\kappa_O x}\left(\sin ky \mathbf{e}_x+\kappa_O \cos ky \mathbf{e}_y\right),\label{eq:b1va}
\end{align}	
with $B_1 = -E_1'/\gamma v$. This perturbation has no electric field component, rendering it magnetostatic.

\begin{figure}
	\centering
	\includegraphics[width = \columnwidth]{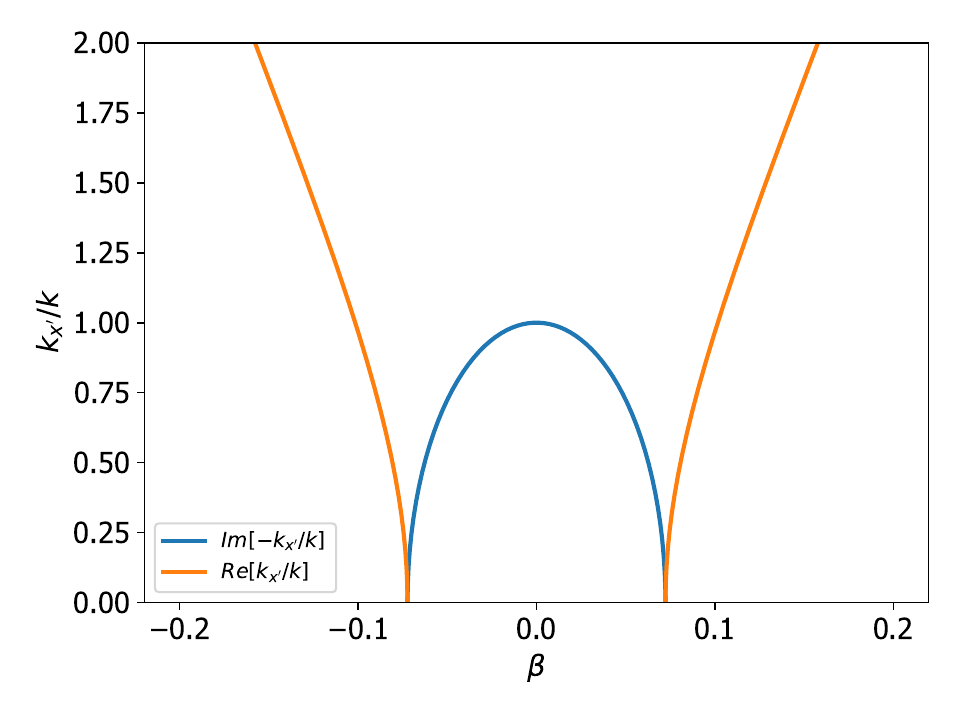}
	\caption{The exact dispersion relation, $k_{x'}$ as a function of $\beta = v/c$, for a quasi-neutral electron-proton plasma, with $n_e =n_p = 10^{20}[m^{-3}]$, $B_0 = 10[T]$, and $k = 40[m^{-1}]$. The cutoffs appears in $\beta =\pm0.07$, and between them the wave is evanescent. Away from the cutoffs, the wave is propagating.}\label{fig:k50ep plasma}
\end{figure}
\begin{figure}
	\includegraphics[width = \columnwidth]{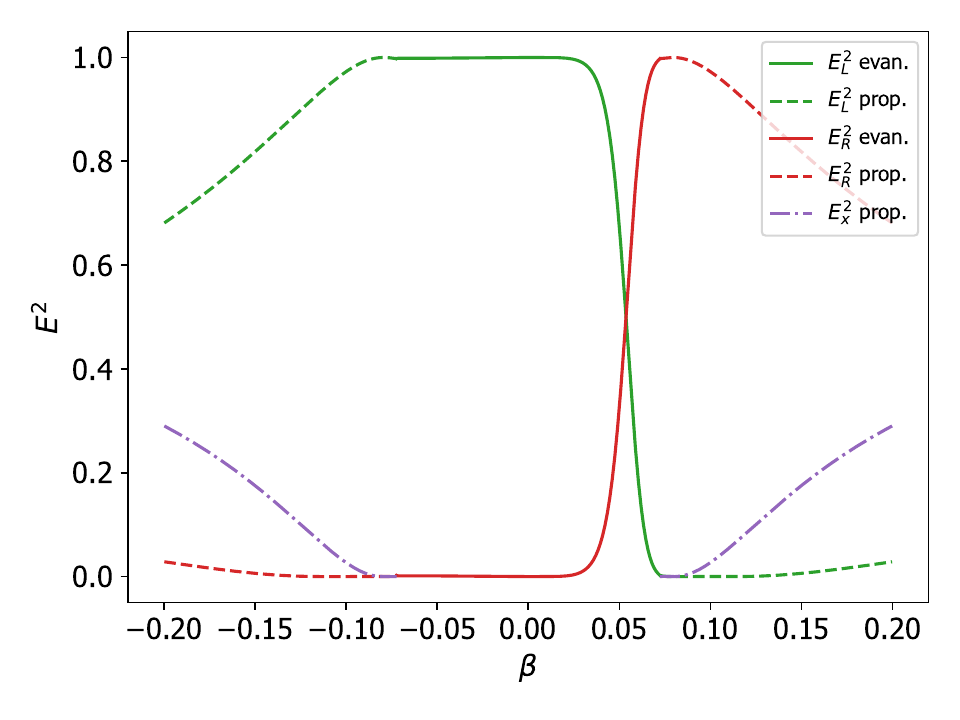}
	\caption{Wave polarization squared in the moving frame, as a function of $\beta$, for the same parameters defined in Figure \ref{fig:k50ep plasma}. Plotted are the squares of the coefficients of the unit vector in the electric field direction. In the evanescent regime, the wave is composed of only right and left polarization, without a phase shift. In the propagating regime, the wave acquires a phase-shift, essentially splitting into three components.} 
\label{fig:k50ep plasma pol}
\end{figure}

\subsubsection{X wave}
The second solution to the dispersion relation is the X wave. The dispersion of the X wave in the frame moving with the flow is
\begin{gather}
	N_{x'}^2=S-\frac{D^2}{S}-\frac{1}{\beta^2},\label{eq: Nx2}
\end{gather} 
which can be positive, negative, or zero. 

When $N_{x'}^2<0$, the X wave is evanescent in $x'$, and
\begin{gather}
	k_{x'} =-ik\sqrt{\beta^2\gamma^2 \left|N_{x'}^2\right|}=-ik \kappa_X,\\
	\kappa_X = \sqrt{1-\beta^2\gamma^2 (S-1)+ \beta^2\gamma^2\frac{D^2}{S}}\label{eq:kappa X}.
\end{gather}
Even for in the evanescent regime, the penetration length is generally much longer than for the O wave. The electric field is oriented in the $x-y$ plane, and its polarization now depend on the value of $\beta$, as well as $S$ and $D$. It is purely imaginary, shifting continuously from circular to elliptic, with 
\begin{gather}
	p=\frac{h_{x'}}{h_{y'}} = i\frac{\gamma\beta^2 D+ \kappa_X}{\gamma\beta^2 S-\gamma},\ \ h_{z'}  =0. \label{eq:pol kappa x}
\end{gather}
In the low-flow and low density limits, $p\rightarrow -i$, which is a left-handed circular polarization.

When $N_{x'}^2>0$ the X wave can propagate in the slab. The polarization becomes complex, with both a real and imaginary components. This is a regime of lesser interest, due to limited application in a cylindrical device. Additionally, for a finite plasma the penetration length of the evanescent regime is sufficient.

The electromagnetic potential of this perturbation in the moving frame is, up to a phase
\begin{gather}
	\mathbf{A}_X' =  \Re\left[-\frac{E_1(z)}{ k\gamma v}\frac{p\mathbf{e}_{x'}+\mathbf{e}_{y'}}{\sqrt{|p|^2+1}}e^{k \kappa_X x'+i k\gamma(y' +vt')}\right]
\end{gather}
and the electromagnetic fields in the limit of $\pdv{E_1'}{z}=0$ are
\begin{gather}
	\mathbf{E}'_{X} = \Re\left[iE_1\frac{p\mathbf{e}_{x'}+\mathbf{e}_{y'}}{\sqrt{|p|^2+1}}e^{k \kappa_X x'+i k\gamma(y' +vt')}\right],\\
		\mathbf{B}'_{X} =- \frac{E_1}{\gamma v}\frac{ \kappa_X+ \gamma \Im[p]}{\sqrt{|p|^2+1}}e^{k \kappa_X x'}\cos (k\gamma(y' +vt'))\mathbf{e}_{z'},
\end{gather}
with $\Re[f]$ being the real part of $f$ and $\Im[f]$ being the imaginary part of $f$.

We can decompose $\mathbf{E}'_{X}$ to the left $\mathbf{e}_{L'} =(\mathbf{e}_{x'}+i\mathbf{e}_{y'})/\sqrt{2} $, right  $\mathbf{e}_{R'} =(\mathbf{e}_{x'}-i\mathbf{e}_{y'})/\sqrt{2} $ and linear polarizations by
\begin{gather}
	E_{L'} = \frac{1}{\sqrt{2}}\frac{\Im[p]-1}{\sqrt{|p|^2+1}},\quad E_{R'} = \frac{1}{\sqrt{2}}\frac{\Im[p]+1}{\sqrt{|p|^2+1}},\label{eq:coeff1}\\
	E_{x'} = \frac{\Re[p]}{\sqrt{|p|^2+1}},\label{eq:coeff2}
\end{gather}
with the linear polarization appearing only in the propagating regime.

An example to the dispersion of an electron-proton quasi-neutral plasma, of fusion-relevant density (if not composition) is plotted in Figure \ref{fig:k50ep plasma}, with the blue curve representing the evanescent regime, and the orange curve representing the propagating regime, from equation (\ref{eq: Nx2}). The polarization is represented by the squares of the coefficients in equations (\ref{eq:coeff1}) and (\ref{eq:coeff2}) in Figure \ref{fig:k50ep plasma pol}. This is a simple plasma, without any resonant interactions in the plotted parameter range. The wave polarization switches from left to right in the case of flow in the positive $y$ direction. 

\begin{figure}
	\centering
	\includegraphics[width = \columnwidth]{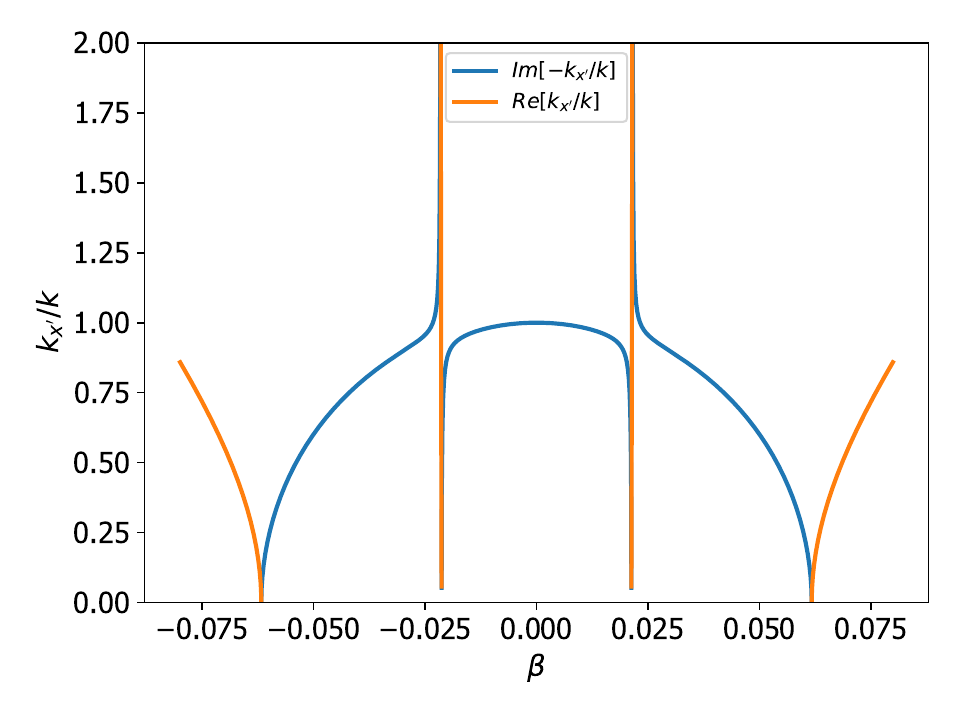}
	\caption{The exact dispersion relation, $k_{x'}$ as a function of $\beta = v/c$, for a quasi-neutral electron-proton-boron11 plasma, with $n_p = 0.5\cdot 10^{20}[m^{-3}]$, $n_{b11} = 0.1\cdot 10^{20}[m^{-3}]$, and $B_0 = 10[T]$, and $k = 100[m^{-1}]$, plotted in solid lines. The wave is evanescent in the regions plotted in blue, and is propagating in the regions plotted in orange. In here, a pair of a new cutoff and a resonance appear around $\beta = \pm 0.0215$. The low-frequency approximation is plotted in dashed lines. }\label{fig:k100epb11 plasma}
\end{figure}

\begin{figure}
	\centering
	\includegraphics[width = \columnwidth]{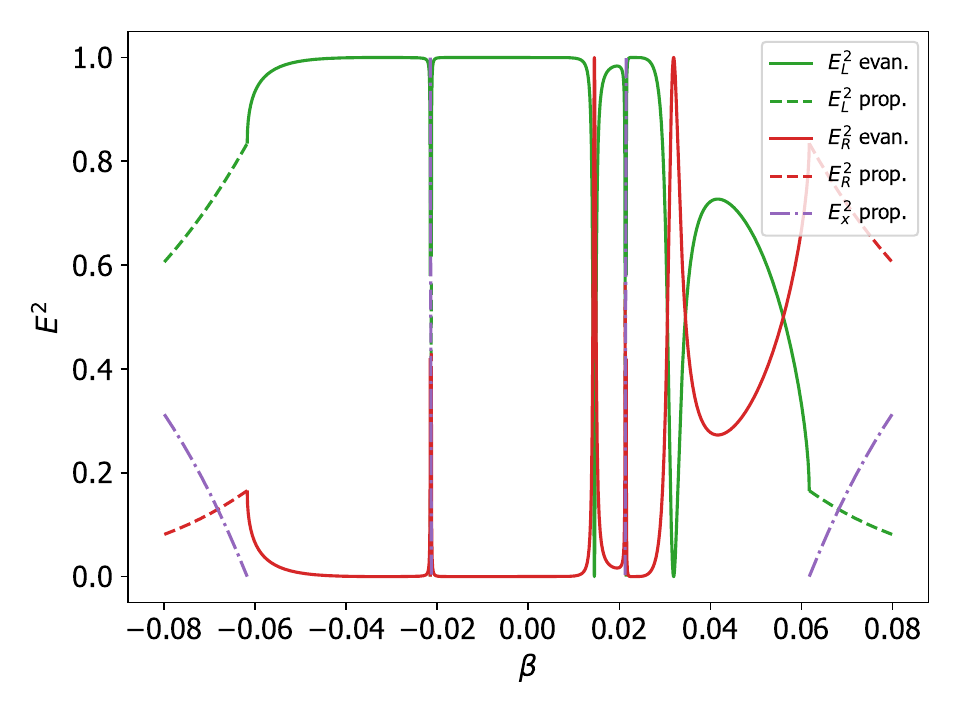}
\caption{Wave polarization squared in the moving frame, as a function of $\beta$, for the same parameters defined in Figure \ref{fig:k100epb11 plasma}. Plotted are the squares of the coefficients of the unit vector in the electric field direction. In the evanescent regime, the wave is composed of only right and left polarization, without a phase shift. In the propagating regime, the wave acquires a phase-shift, essentially splitting into three components. See Figure \ref{fig:k100RenormE1} for the ponderomotive potentials.} \label{fig:k100epb11 plasma pol}
\end{figure}

In contrast, a second example - the dispersion of a electron-proton-boron11 quasi-neutral plasma of the same electron density is plotted in Figure \ref{fig:k100epb11 plasma}, and the polarization in Figure \ref{fig:k100epb11 plasma pol}. In this case, the boron cyclotron resonance is visible as near vertical features in Figure \ref{fig:k100epb11 plasma}.The wave polarization varies much more rapidly in this case.


The Lorentz transform of these fields to the lab frame gives, 
\begin{gather}
	\mathbf{E}_{X}= \frac{E_1 e^{k\kappa_X x}}{\sqrt{\left|p\right|^2+1}} \left( \kappa_X\sin ky\mathbf{e}_x+\cos ky\mathbf{e}_y\right),\label{eq:electric component X wave}\\
	\mathbf{B}_{X}= -\frac{E_1}{v}\frac{ \kappa_X+\Im[p]/\gamma}{\sqrt{\left|p\right|^2+1}}e^{k\kappa_X x}\sin ky\mathbf{e}_{z}.\label{eq:magnetic component X wave}
\end{gather}
In the low-flow rate or low density limits, this perturbation is dominated by its electric component in equation (\ref{eq:electric component X wave}) over its magnetic component. While the magnetic component in equation (\ref{eq:magnetic component X wave}) is never exactly zero, this is nearly an electrostatic perturbation.

\subsection{Cylindrical geometry}
The situation in the cylinder can be more involved than in the slab. We start again from the Hamiltonian given in equation (\ref{eq:Ham general}), this time with potentials given by
\begin{eqnarray}
    \Phi &=&  \frac{1}{2}r^2B_0\omega_{E\times B} = \frac{1}{2}B_0\omega_{E\times B}(x^2+y^2),\label{eq:Phi0}\\
    \mathbf{A} &=& \frac{1}{2}rB_0\mathbf{e}_\phi = \frac{1}{2}(x\mathbf{e}_y-y\mathbf{e}_x)B_0, \label{eq:A0}
\end{eqnarray}
which generate the electric and magnetic fields $\mathbf{E} = -rB_0\omega_{E\times B}\mathbf{e}_r$ and $\mathbf{B}=B_0 \mathbf{e}_z$. The frequency $\omega_{E\times B}$ is a parameter defining the strength of the electric field which generate the rotation. 

Using the cyclotron frequency $\Omega$, we define the effective cyclotron frequency in this geometry~\cite{brillouinTheoremLarmorIts1945,davidsonPhysicsNonneutralPlasmas1990}
\begin{gather}
	\Omega_B =sign(\Omega)\sqrt{\Omega^{2}+4\omega_{E\times B} \Omega}.
\end{gather}
Particles are confined in these fields as long as  $\Omega_B\in \mathbb{R}$, or $\omega_{E\times B}/\Omega > -1/4$.

Using the generating function $F_2 = F_2(p_x, y,\theta,\varphi)$,
\begin{multline}
    F_2 = \frac{1}{8 m\Omega_B}(2 p_x-m \Omega_B y )^2\cot (\theta_1 ) \\
        +\frac{1}{8m\Omega_B} (2 p_x+ m\Omega_B y)^2\cot (\theta_2),
\end{multline}
for the canonical transformation of the $(x,y,p_x,p_y)$ coordinates to the $(\theta_1,\theta_2,I_1,I_2)$ coordinates such that the gyrocenter radius is $R_G = \sqrt{2I_1 / m\Omega_B}$ and the gyroradius is $\rho = \sqrt{2I_2 / m\Omega_B}$, and 
\begin{gather}
    x =R_G\cos \theta_1 -\rho\cos\theta_2,\quad 
    y=R_G\sin  \theta_1 +\rho\sin \theta_2 ,\nonumber\\
    p_{x} =-R_G\sin  \theta_1 +\rho\sin \theta_2 ,\quad
    p_{y}=R_G\cos  \theta_1 +\rho\cos \theta_2.\label{eq:cylinder action angle}
\end{gather}
A second canonical transformation step using $F_3=F_3(I_1,I_2,\phi,\theta)$, transforms $I_1$ and $I_2$ into the first adiabatic invariant $\mu$ and the canonical angular momentum $p_\phi$.
\begin{align}
    &F_3 = -(I_1 -I_2)\phi- I_2\theta,\\
    &p_\phi =I_1 -I_2, &\phi& = \theta_1\nonumber \\
    &\mu_0 =I_2,  &\theta& = \theta_1+\theta_2. \label{eq:cylinder action angle2}
\end{align}

At the end of these transformations, the Hamiltonian becomes
\begin{gather}
	H = \frac{p_z^2}{2m}+\Omega_B\mu+\omega_{rot}p_\phi,
\end{gather}
with the rotation frequency around the device being
\begin{gather}
	\omega_{rot} = \frac{1}{2}(\Omega_B-\Omega) = \omega_{E\times B}+\frac{1}{2}\left(\Omega_B - \sqrt{\Omega_B^2+4\omega_{E\times B}^2}\right).
\end{gather}
In the event where $\omega_{E\times B}$ is a constant, but $B_0$ is a function of z, the rotation frequency $\omega_{rot}$ is \textit{not} a constant. This effect generates the centrifugal potential in the presence of isorotating drift surfaces. 

We expect to be in the limit of $\omega_{E\times B}\ll \Omega_B, \Omega$. 

The main consequence of the solution to the motion in crossed fields in the cylinder is that the flow velocity around the device $r \omega_{rot}$, is no longer the same for all species as the flow velocity was in the slab. As such, the same trick of using a frame transformation would not work in the cylinder, for the reason that there is no frame in which the plasma is stationary. In addition, the motion around a cylinder would generate a non-inertial frame transformation.  

Additionally, this difference in flow velocities between species can be observed as an azimuthal current in the plasma. This current would render the magnetic field non uniform, $B_0=B_0(r)$, which complicates the solution further.  

However, we did take a useful piece of information from the slab - in the tenuous plasma limit and in the low flow limit, a perturbation to the plasma is close to vacuum fields.

In the vacuum limit, we can apply either an electric or magnetic multipole fields, which are analogous to the O wave and the X wave in the slab
\begin{gather}
	\mathbf{A}_O = -B_1\frac{R}{n}
        \left(\frac{r}{R}\right)^{n}\cos n\phi \mathbf{e}_{z},\ \ r< R,\\
    \mathbf{B}_O = B_1
        \left(\frac{r}{R}\right)^{n-1}\left(\sin\left(n\phi\right)\mathbf{e}_r+\cos\left(n\phi\right)\mathbf{e}_\phi\right),\ \ r< R,
\end{gather}
or
\begin{gather}
	\Phi_X = E_1\frac{R}{n}
        \left(\frac{r}{R}\right)^{n}\sin\left(n\phi\right),\ \ r< R,\\
    \mathbf{E}_X = E_1
        \left(\frac{r}{R}\right)^{n-1}\left(\sin\left(n\phi\right)\boldsymbol{e}_r+\cos\left(n\phi\right)\mathbf{e}_\phi\right),\ \ r< R.
\end{gather}
In the cylinder the wave vector component $k_y$ becomes $n/R$ when $n\in \mathbb{N}$.

\section{Ponderomotive potentials}\label{sec: ponderomotive potentials}
In order to derive the ponderomotive potential for a perturbation, we look at the contribution of the perturbation to the Hamiltonian in action-angle coordinates. This can be achieved by taking the perturbation Hamiltonian to be 
\begin{gather}
	H_1 = -\frac{\mathbf{p}\cdot e\mathbf{A}_1}{m}+\frac{e^2A_1^2}{2m}+e\Phi_1
\end{gather}
with the appropriate $\mathbf{A}_1$ and $\Phi_1$ for the perturbation. 

For a ponderomotive potential, we require a separation of time scales. In the slab, this is
\begin{gather}
	\left|\pdv{H_1}{z}v_z\right|\ll \left|\pdv{H_1}{Y}v\right|,
\end{gather}
and in the cylinder, 
\begin{gather}
	\left|\pdv{H_1}{z}v_z\right|\ll \left|\pdv{H_1}{\phi}\omega_{rot}\right|.
\end{gather}

Another requirement is for the beat period between the gyro motion harmonics and the interaction with the perturbations to be smooth. In the slab
\begin{gather}
	\forall \ell: \frac{v_z}{\ell\Omega-vk}\frac{1}{L} \ll1,
\end{gather}
and in the cylinder
\begin{gather}
	\forall\ 0\le\ell\le n: \frac{v_z}{\ell\Omega_B-n\omega_{rot}}\frac{1}{L} \ll1,\\
	\forall\ 0\le\ell\le n: \frac{v_z}{\ell\Omega_B-2n\omega_{rot}}\frac{1}{L} \ll1,
\end{gather}

with $L$ being the ramp-up length scale of the perturbation.

\subsection{Magnetostatic perturbation}
For a magnetostatic perturbation with $\mathbf{A}_O\parallel \mathbf{e}_z$, the perturbation Hamiltonian consists of a term $-p_z e A_O/m$ which does not contribute to the ponderomotive potential, but generates a mass modification term in the dynamics, and a term ${A}_O^2/2m$. This second term generates a ponderomotive potential term that is to leading order in the perturbation amplitude simply $\langle{A}_O^2/2m\rangle$, with the triangular brackets denoting an average over the oscillations. 

In order to take into account the leading order effects for small but nonzero gyroradius, the two term ${A}_O^2/2m$ in the Hamiltonian has to be Fourier expanded into a series in $\theta$ after transformation to the action-angle variables using equations (\ref{eq:slab action angle}) and (\ref{eq:cylinder action angle}), (\ref{eq:cylinder action angle2}). 
The ponderomotive potential for the magnetostatic perturbation in the slab is 
\begin{gather}
	\Phi_\mathrm{pond,O} = \frac{B_1^2 e^{2k\kappa_O X}}{4mk^2}I_0(2\kappa_O\rho),
\end{gather}
with $I_0$ being the modified Bessel function of the first kind of order 0.
For the cylinder, the expression is possibly simpler,
\begin{gather}
	\Phi_\mathrm{pond,O} = \frac{B_1^2}{4m}\frac{R^2}{n^2}\left(\frac{R_G^2+\rho^2}{R^2}\right)^{n}.
\end{gather}

This ponderomotive potential is always positive, i.e., repulsive, and repels particles from regions of high $B_1$.

It is interesting to note that the Fourier expansion of the Hamiltonian in the slab case is a sum with infinite terms, while in the cylinder the sum ends up having finitely many terms. 

\begin{figure*}
	\centering
	\includegraphics[width = \textwidth]{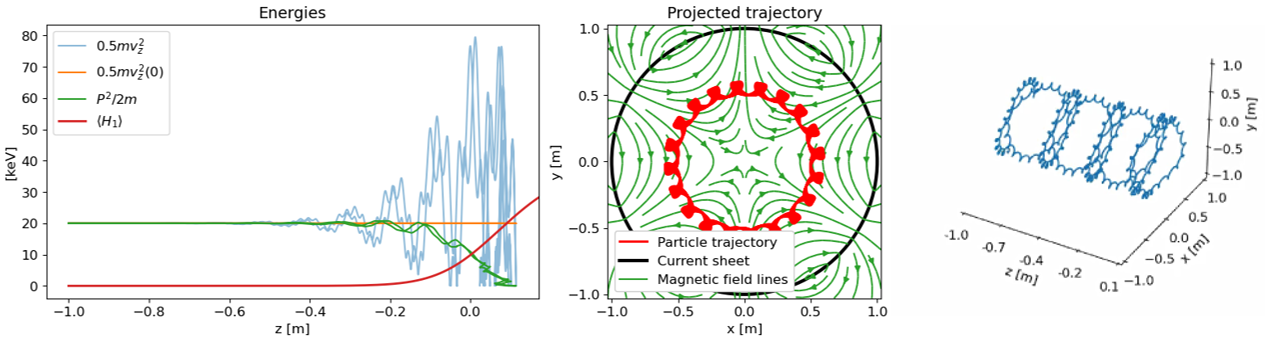}
	\caption{Particle trajectory interacting with a magnetostatic ponderomotive barrier. In the Left figure: energies as a function of axial position. In blue, the kinetic energy in the z direction. In orange, the initial kinetic energy in the z direction. In green, the energy in the ballistic motion in the z direction. In red, the ponderomotive barrier. In the Middle figure: In red, a projection of the particle trajectory on the x-y plane. In green, the perturbation magnetic field. In the Right figure: a 3D render of the particle trajectory. \label{fig: trajectory}}
\end{figure*}

The ponderomotive potential is generated in this case by oscillations along the z direction. This is due to the unbalanced force generated by the $v_y\mathbf{e}_y\times B_x \mathbf{e}_x = -v_y B_1 \cos (kv_y t) \mathbf{e}_z$. This can be observed in Figure \ref{fig: trajectory}, where the kinetic energy in the z degree of freedom rises as the particle interacts with the perturbation potential, while the energy in the ballistic degree of freedom decreases in approximately the same amount as the ponderomotive potential. This figure was obtained by a full-orbit numerical simulation using a second-order generalization of Boris' method \cite{zenitaniBorisSolverParticleincell2018,boris1970relativistic,qinWhyBorisAlgorithm2013}, using the LOOPP code which was used in several of our previous publications \citep{ochsNonresonantDiffusionAlpha2021,ochsPonderomotiveRecoilElectromagnetic2023,rubinGuidingCentreMotion2023, rubinMagnetostaticPonderomotivePotential2023,rubinFlowingPlasmaRearrangement2024a}.

\subsection{Electrostatic perturbation}
For an electrostatic perturbation with $\mathbf{E}'_X\perp \mathbf{e}_{z'}$, the perturbation Hamiltonian consists of a term $-\mathbf{p}\cdot \mathbf{A}_X/m$ and  ${A}_X^2/2m$, both of which contribute to the ponderomotive potential. At leading order in the perturbation amplitude, the ponderomotive potential is still the average over the oscillations of the two terms. 

The ponderomotive potential in the small gyroradius limit for a perturbation with a polarization in the x-y plane is given by
\begin{gather}
\Phi_\mathrm{pond,X}=\frac{e^2}{4m\omega_\mathrm{wave}}\left(\frac{E_L^2}{\omega_\mathrm{wave}+\Omega}+\frac{E_R^2}{\omega_\mathrm{wave}-\Omega}\right),\label{eq: pond pot X}
\end{gather}
with the coefficients determined by the polarization as in the slab in the evanescent regime, and
\begin{gather}
	\omega_\mathrm{wave} = \frac{kE_0}{B_0}.
\end{gather} 

This potential can be used to attract the plasma to a specific axial region, or repel the plasma from it such as in an end plug. In order to do so effectively, we must utilize the cyclotron resonance, in order to attract or repel at least one species of ions, while not affecting the electrons as strongly. 
Trying to use the $\omega_\mathrm{wave}\approx0$ pole yields for all species
\begin{gather}
\Phi_\mathrm{pond,X}\approx\frac{e}{4\omega_\mathrm{wave}}\frac{E_L^2-E_R^2}{B_0},
\end{gather}
which is proportional to the particle charge, and has opposite signs for electrons and ions. It has the same magnitude for electrons and singly-ionized ions. In an electron-proton plasma, use of this pole would generate no net confinement or deconfinement. 

If instead we would attempt to use the first cyclotron resonance, $\omega_\mathrm{wave}\approx \Omega$, such as $\omega_\mathrm{wave} = \alpha \Omega_i$ with $\alpha \approx -1$, with $E_R=0$, 
\begin{gather}
\Phi_\mathrm{pond,X,ions}\approx\frac{m_i}{4 \alpha B_0^2}\frac{E_L^2}{\alpha+1},\label{eq:pond ions}\\
\Phi_\mathrm{pond,X,electrons}\approx-\frac{m_i}{ Z_i}\frac{E_L^2}{4\alpha B_0^2}.
\end{gather}
With $Z_i$ being the ion charge number. The sign of equation (\ref{eq:pond ions}) depends on the sign of $\alpha+1$, and the ratio of the poderomotive potentials for the different species is 
\begin{gather}
\frac{\Phi_\mathrm{pond,X,ions}}{\Phi_\mathrm{pond,X,electrons}}\approx-\frac{Z_i}{\alpha+1}.
\end{gather}
which can be quite large for $\alpha$ close enough to $-1$.

\begin{figure}
	\centering
	\includegraphics[width = \columnwidth]{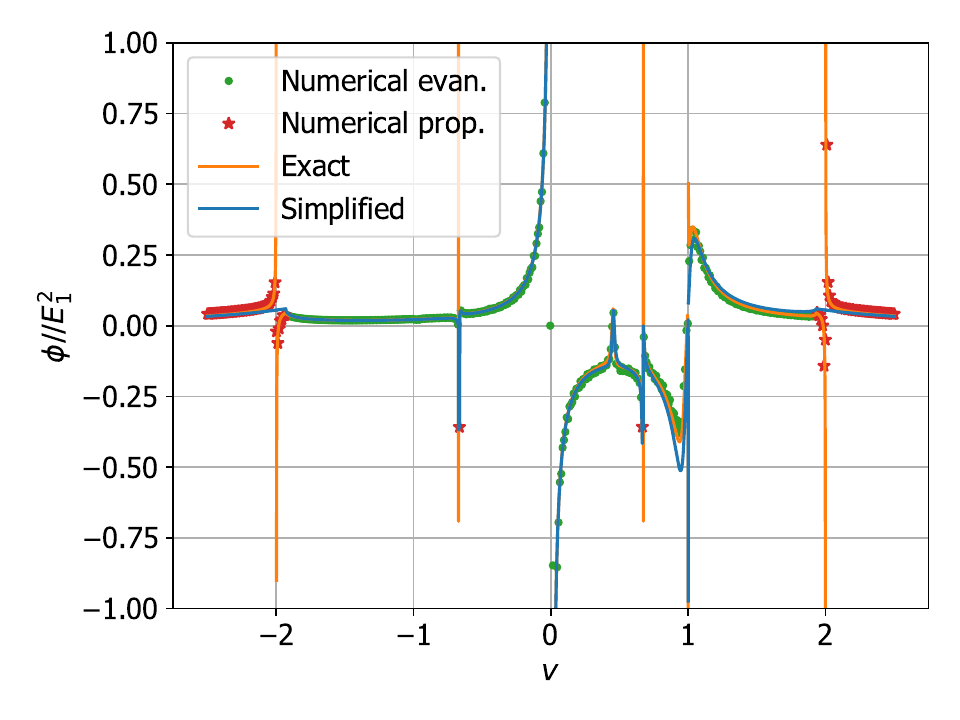}
	\caption{Numerical evaluation of the ponderomotive potential (sans the $E_1^2$ amplitude) for protons in a proton-boron11 plasma, with $n_p = 0.5\cdot10^{20}[m^{-3}]$ and $n_{B11}=0.1\cdot10^{20}[m^{-3}]$, $k=100[m^{-1}]$, $B_0=10[T]$ and $\rho= 0.3$. Notice we see here the 2nd resonances at $v=1,\pm 2$, and no resonance at $v=-1$. The green markers indicate a regime where the perturbation is evanescent, and the red markers indicate a propagating perturbation. Most of the ponderomotive potential profile can be explained by the variations in the wave polarization, which is represented in the simplified expression. Figure taken from Ref~[\onlinecite{rubinFlowingPlasmaRearrangement2024a}].}\label{fig:k100RenormE1}
\end{figure}

Due to the dependence of the polarization on the flow velocity, i.e. the wave frequency, the ponderomotive potential can have complex features if the polarization has these complex features, as can be seen in Figure \ref{fig:k100RenormE1}. In this figure, we use the same electron-proton-boron11 plasma as in Figures \ref{fig:k100epb11 plasma} and \ref{fig:k100epb11 plasma pol}, and a perturbation with  $k=100[m^{-1}]$. The relativistic $\beta = v/c$ is plotted on top horizontal axis, while the dimensionless velocity $vk/\Omega$ is plotted on the bottom axis. The red and green dots are the result of the numerical simulation, whereas the solid blue and orange lines are the result of the analytic expression. It is evident that the ponderomotive potential generated by this perturbation can be positive or negative even taking the changing polarization into account. 

The polarization of the perturbation is changing from left to right exactly when the at the proton cyclotron resonance at $v=1$, making the pole less pronounced than the $v=0$ pole. This limits the amplification of the ponderomotive effect for being close to resonance.

\section{Conclusion}

	The ponderomotive effect can be used to generate effective potential barriers which are particularly useful in linear plasma confinement devices and mass separators. We have shown this effect can be generated using static electromagnetic fields that generate a flow through a perturbation. This is done through a Doppler shift (or Lorentz boost) of the static perturbation into a time dependent wave in the frame moving with the flow. 
	
	For particles in the non-relativistic limit in both flowing and lab systems, the ponderomotive effect due to interaction with a static perturbation depends on the polarization of electric field component of the perturbation in the system moving with the flow. Flute like $k_\parallel=0$ perturbations in the fluid limit can exist in two modes, corresponding to the O wave and the X wave. In the moving frame, the electric field of the O wave is polarized in the direction of the static magnetic field, while the electric field of the X wave is polarized perpendicular to it.
	
	The Lorentz boost of the O wave to the lab frame yields a pure magnetic field perturbation perpendicular to the static magnetic field of the magnetic mirror machine. This perturbation penetrates the plasma only in regions of low density, which may occur inside the mirror throats. The ponderomotive effect of this configuration is a positive, i.e., repulsive potential barrier regardless of the sign of the particle charge. It could be used as an end plug, to confine the tail end of the particle population that would have otherwise escape the mirror, after overcoming the magnetic mirror and centrifugal potentials. 
	
	The Lorentz boost of the X wave to the lab frame yields a near electrostatic perturbation in the low flow regime. Low flow requires less recirculating power and lower static electric fields to generate this flow. This perturbation penetrates the plasma well, and can generate either a positive or negative potentials. In the low flow regime, the X wave polarization is not affected by the flow, and the ponderomotive potential appears as in equation (\ref{eq: pond pot X}). 
	
	Utilization of the X wave-like perturbation to attract or repel the plasma must utilize a Doppler-shifted frequency close to a cyclotron resonance, the ion one being more convenient. Using a frequency near zero causes a potential with opposite signs for electrons and ions, separating the electrons from the rest of the plasma. 
	
	Considering the change in polarization from left to right, using a Doppler-shifted frequency close to the ion cyclotron frequency yields a reduced effect, as is visible in Figure \ref{fig:k100RenormE1}, where the pole near $v=1$ is less significant than the pole around $v=0$.  This limits the magnitude of the ponderomotive interaction, but allows for the repulsion or attraction of the plasma as a whole, depending on whether or not $v$ is smaller or larger than 1.

\section*{Acknowledgments}
The authors would like to thank Ian Ochs, Elijah Kolmes, and Alex Glasser for useful discussions. This work was supported by ARPA-E Grant No. DE-AR0001554 {and NSF Grant PHY-2308829.}

\section*{References}
\input{popwriteup.bbl}

\end{document}

%% file: popwriteup.bbl
%